\documentclass[amsmath, amssymb, aps, pra, twocolumn,floatfix,nofootinbib]{revtex4}
\usepackage[utf8]{inputenc}
\usepackage[urlcolor=blue, colorlinks=true,citecolor=blue]{hyperref}
\usepackage{enumitem}

\usepackage{amsmath}
\usepackage{amsmath, amssymb}
\usepackage{textcomp, gensymb}
\usepackage{tabularx}
\usepackage{siunitx}
\usepackage{bm}
\usepackage{wrapfig}
\usepackage{graphicx}
\usepackage{float}
\usepackage{subfigure}
\usepackage{amssymb}
\usepackage{bbold}
\usepackage{color}
\usepackage{ulem}
\usepackage{qcircuit}

\begin{document}

\title{Multi-stream physics hybrid networks for solving Navier-Stokes equations}

\author{Aleksandr Sedykh}
\author{Tatjana Protasevich}
\author{Mikhail Surmach}
\author{Arsenii~Senokosov}
\author{Matvei Anoshin}
\author{Asel Sagingalieva}
\author{Alexey Melnikov}
\thanks{Corresponding author: \texttt{alexey@melnikov.info}}
\affiliation{Terra Quantum AG, 9000 St. Gallen, Switzerland}


\begin{abstract}
Understanding and solving fluid dynamics equations efficiently remains a fundamental challenge in computational physics. Traditional numerical solvers and physics-informed neural networks struggle to capture the full range of frequency components in partial differential equation solutions, limiting their accuracy and efficiency. Here, we propose the Multi-stream Physics Hybrid Network, a novel neural architecture that integrates quantum and classical layers in parallel to improve the accuracy of solving fluid dynamics equations, namely ``Kovasznay flow'' problem. This approach decomposes the solution into separate frequency components, each predicted by independent Parallel Hybrid Networks, simplifying the training process and enhancing performance. We evaluated the proposed model against a comparable classical neural network, the Multi-stream Physics Classical Network, in both data-driven and physics-driven scenarios. Our results show that the Multi-stream Physics Hybrid Network achieves a reduction in root mean square error by 36\% for velocity components and 41\% for pressure prediction compared to the classical model, while using 24\% fewer trainable parameters. These findings highlight the potential of hybrid quantum-classical architectures for advancing computational fluid dynamics.
\end{abstract}

\maketitle

\section{Introduction}
Computational fluid dynamics (CFD) is a branch of fluid dynamics that develops a range of techniques to analyze and solve problems involving the dynamics of fluid flows. This discipline is widely used in many fields of science, including aerodynamics, weather prediction, engine design, and biological engineering. The vast majority of all fluid dynamics problems revolves around the Navier-Stokes equations, which determine the motion of Newtonian viscous fluids.

\begin{equation}\label{eq:ns}
    \frac{\partial \bm{v}}{\partial t} = -(\bm{v} \cdot \nabla) \bm{v} + \nu \Delta \bm{v} - \frac{1}{\rho} \nabla p + \bm{f},
\end{equation}
where $\nabla$ is nabla operator, $\Delta$ is Laplace operator, $t$ is time, $\nu$ is kinematic viscosity, $\rho$ is fluid density, $p$ is pressure, $\bm{v}$ is velocity 
vector, $\bm{f}$ is external forces. To solve Navier-Stokes 
equations means to find $p$ and $\bm{v}$ as functions of 
coordinate and time: $p(\bm{r}, t)$ and $\bm{v}(\bm{r}, t)$ \cite{cfd_review, Anderson1995ComputationalFD}. 

Computational fluid dynamics deals with the numerical solution of these equations, since analytical solutions are known only for some special cases (one of which will be considered in this paper). To obtain these numerical solutions, the so-called ``solvers''--computer programs that use the finite element method (or any other numerical method) to approximate the Navier-Stokes equations~\cite{openfoam, ansys}. The solvers partition the fluid volume into a large number of cells~\cite{maric2020unstructured} where it is easier to get an approximate solution, and then combine the solution from all the cells to obtain the velocity and pressure distribution across the entire geometry~\cite{Mari2013voFoamA}. Although this is a rather crude explanation of the workflow of these programs, solvers have one major drawback: since the solution is obtained numerically, any change in the parameters of the original problem leads to an inevitable reset of the entire simulation, which can take quite a long time~\cite{cai2021physics}. 

The proposed solution to this problem is to use a neural network as a solver. Neural networks are universal function approximators, that is, with a sufficient number of neurons (parameters), they can approximate any function as accurately as desired. This gives a theoretical justification of the possibility to learn the solution of the Navier-Stokes equations. Moreover, this solution can be learned immediately in a large range of values of some parameter $\lambda$ (for example, kinematic viscosity $\nu$) by passing it to the neural network as another coordinate together with $\bm{r}$ and $t$. Thus, after training the neural network, an integer parameterized set of solutions can be obtained. This can be useful, for example, in optimal parameterization problems.

Recent theoretical and practical studies show that the number of parameters required by a neural network to approximate the vector function $\bm{\varphi}: \mathbb{R}^d \rightarrow \mathbb{R}$ depends on $d$ polynomially~\cite{hutzenthaler2020proof, grohs2023proof}.
At the same time, to solve some problem of dimension $d$ using the finite difference method, it will be necessary to calculate the differential equation at points $N^d$, where $N$ is the number of points along one of the axes determined by the step size of the method; that is, the complexity of the problem in this case grows exponentially. It should be concluded that the application of neural networks in problems of high dimensionality can be well justified.

This paper explores neural networks collectively called ``Physics-Informed Neural Networks'', hereafter PINNs, first introduced in the \cite{raissi2019physics}. These neural networks can be used to solve any parameterized differential equation; they do not require linearization or discretization~\cite{cai2021physics}. One only needs to specify the model architecture (like multi-layered perceptron or Multi-stream Physics Hybrid Network introduced in this paper) and choose a suitable loss function. The ``physics'' part in the name refers to the fact that these neural networks use physical laws (differential equations describing a particular problem) for training, rather than a ready-made solution obtained, for example, by a solver (this approach is called ``data-driven''). Although it is possible to use these two approaches together and create a combined model~\cite{li2024physics}, in this paper we focus on physics-driven models. When trained, such models minimize the error of the differential equation. Thus, as the training progresses, the neural network satisfies it better and better. 

Machine learning can be greatly improved by using quantum technologies. In \cite{gaitan2020finding} quantum computing is used within a similar problem. The performance of existing machine learning models is limited by high computational resource requirements. Quantum computing can improve the learning process of classical models, allowing for better accuracy in predicting the target function with fewer iterations~\cite{dunjko2018machine, qml_review_2023, Neven2012QBoostLS, rebentrost2014quantum, saggio2021experimental, kordzanganeh2023parallel, kordzanganeh2023benchmarking}. In many industrial and scientific fields, such as pharmaceutical~\cite{sagingalieva2023hybrid, gircha2021training}, aerospace~\cite{rainjonneau2023quantum}, automotive~\cite{sagingalieva2023hyperparameter}, and financial~\cite{alcazar2020classical, coyle2021quantum, pistoia2021quantum, emmanoulopoulos2022quantum, raj2023quantum} quantum technologies can provide improvements to existing classical methods. Many traditional machine learning tasks such as image processing~\cite{senokosov2023quantum, li2022image, naumov2025tensor, lusnig2024hybrid, riaz2023accurate, sagingalieva2025hybrid}, time series forecasting~\cite{kurkin2025forecasting,sagingalieva2025photovoltaic,lee2025predictive}, and natural language processing~\cite{hong2022qspeech, lorenz2023qnlp, coecke2020foundations, meichanetzidis2020grammar} have already demonstrated the broad application prospects of quantum methods. The goal of this work is to explore the feasibility of applying quantum machine learning to the new and emerging field of physical modeling using neural networks.

The great success of neural networks is due in large part to automatic differentiation technology~\cite{autodiffPaper}, which allows us to efficiently read the gradients of the loss function over the model parameters. In our problem, automatic differentiation is also needed to compute the differential equation inside the loss function. Moreover, unlike solvers, in this case the residual value of the differential equation at each spatial coordinate will be calculated exactly, eliminating the need to use approximate differentiation techniques. It follows that PINNs do not require any special discretization of the problem geometry. However, we still need some points $(\bm{r}, t; \lambda)$ in order to compute the differential equation at them, count its error, and minimize this error during training. In order to obtain a solution close to the exact solution using a PINN, it is important that the model has a sufficiently high expressivity (ability to approximate a large number of functions). Fortunately, expressivity is a well-known advantage of quantum computers~\cite{quantum_expressivity, anoshin2024hybrid}. Moreover, quantum circuits with particular structure are differentiable, which allows them to be used in this problem.

Here, we propose a new PINN model architecture -- Multi-stream Physics Hybrid Network (MPHN). This approach uses several Parallel Hybrid Networks (PHNs) for predicting different components of the solution vector. PHN itself is a network consisting of two parts -- quantum and classical layers. Such modular architecture allows for both flexibility and training simplicity. 

We evaluated the proposed MPHN on the ``Kovasznay flow'' problem of modelling laminar fluid flow behind a two-dimensional grid. This problem has an exact solution obtained by Leslie S. G. Kovasznay \cite{kovasznay1948laminar}. The solution accuracy of the model can be correctly estimated due to the fact that the problem has an exact solution. In problems where there is no exact solution, it is rather difficult to make such an estimation and usually one has to resort to experiments (e.g., in wind tunnels).

\section{Kovasznay flow problem formulation}
We will be solving the Navier-Stokes equations in rectangular domain $\Omega = [-0.5, 1.0] \times [-0.5, 1.5]$. Kovasznay flow model describes the fluid flow behind two dimensional grid.
The flow is laminar and is governed by $2$D Navier-Stokes equations:
\begin{align}\label{eq:ns_kovasznay}
\begin{split}
    v_x \frac{\partial v_x}{\partial x} + v_y \frac{\partial v_x}{\partial y} &= - \frac{\partial p}{\partial x} + \frac{1}{\text{Re}} \left( \frac{\partial^2 v_x}{\partial x^2} + \frac{\partial^2 v_x}{\partial y^2} \right), \\
    v_x \frac{\partial v_y}{\partial x} + v_y \frac{\partial v_y}{\partial y} &= - \frac{\partial p}{\partial y} + \frac{1}{\text{Re}} \left( \frac{\partial^2 v_y}{\partial x^2} + \frac{\partial^2 v_y}{\partial y^2} \right),
\end{split}
\end{align}
where $v_x, v_y$ are velocity projections, $p$ is pressure, $\text{Re}$ is the Reynolds number.

The analytical solution of these equations was discovered by Leslie S. G. Kovasznay in 1948 \cite{kovasznay1948laminar}.
\begin{align}\label{eq:exact}
\begin{split}
    v_x^\text{e} &= 1 - e^{\lambda x} \cos (2 \pi y), \\
    v_y^\text{e} &= \frac{\lambda}{2 \pi} e^{\lambda x} \cos (2 \pi x), \\
    p^\text{e} &= \frac{1}{2} (1 - e^{2 \lambda x}),
\end{split}
\end{align}
with boundary conditions
\begin{align}\label{eq:bc}
\begin{split}
    v_x &= v_x^\text{e}, \quad x \in \partial \Omega, \\
    v_y &= v_y^\text{e}, \quad x \in \partial \Omega, \\
    p   &= p^\text{e},   \quad x = 1,
\end{split}
\end{align}
and parameter
\begin{equation}
    \lambda = \frac{1}{2 \nu} - \sqrt{\frac{1}{4 \nu} + 4 \pi^2}.
\end{equation}
Thus, the velocity boundary conditions are set on the entire boundary of the region $\partial \Omega$, and the pressure boundary conditions are set on the right wall $x=1$. For all subsequent simulations we will be using Reynolds number $\text{Re} = 20$ and kinematic viscosity $\nu = 1 / \text{Re} = 0.05$.

\section{Physics-informed neural networks}
PINNs were first introduced in \cite{raissi2019physics}. The idea behind these models is to use a neural network -- usually a feedforward neural network, such as a multilayer perceptron -- as a solution function of a differential equation. Consider an abstract parameterized partial derivative equation (hereafter, PDE):
\begin{equation}\label{eq:abs_pde}
    \mathcal{D}[f(\bm{r}, t); \lambda] = 0,
\end{equation}
where $\bm{r} \in \Omega$ is a coordinate vector, $\Omega \subset \mathbb{R}^d$ is problem definition domain (with dimension $d$), $t \in \mathbb{R}$ is time, $\mathcal{D}$ is nonlinear differential operator parameterized by $\lambda$ parameters, $f(\bm{r}, t)$ is solution function.

Consider a neural network $u(\bm{r}, t)$ that takes coordinates and time as input and outputs some real number (e.g., the pressure of a liquid at a certain point, at a certain moment in time). We can compute the function $u(\bm{r}, t)$ at any point in the problem definition domain by making a so-called forward pass in the neural network, and we can also compute its derivatives of any order $\partial_t^n u(\bm r, t)$, $\partial_{\bm r}^n u(\bm r, t)$ by making a backward pass \cite{rumelhart1986learning}. It turns out that one can simply replace $f(\bm{r}, t) \rightarrow u(\bm{r}, t)$ and attempt to learn the solution to the PDE using standard gradient optimization techniques (e.g., gradient descent).
The advantage of this approach to solving PDEs is the ability to compute exact derivatives of neural networks (standard solvers are forced to use approximate diffusion techniques, e.g., using difference schemes), and the ability of neural networks to approximate complex functions~\cite{autodiffPaper, Hornik1989MultilayerFN, sedykh2024hybrid}.

The loss function that PINN must minimize consists of two summands
\begin{equation}\label{eq:physics_driven_loss}
    \mathcal{L} = \mathcal{L}_\text{PDE} + \mathcal{L}_\text{BC},
\end{equation}
where $\mathcal{L}_\text{BC}$ is responsible for satisfying the boundary conditions and $\mathcal{L}_\text{PDE}$ is responsible for satisfying the PDE itself.

Let us define $\mathcal{L}_\text{BC}$. Consider a boundary condition of Dirichlet form~\cite{greenshields2022notes} for some component in the solution:
\begin{equation}
    u(\bm{r}, t)|_{\bm{r} \in B} = u_0(\bm{r}, t),
\end{equation}
where $u_0(\bm{r}, t)$ specifies the value of the function at the boundary and $B \in 
\mathcal{R}^d$ defines, in fact, this boundary. If $u(\bm{r}, t)$ is a neural network, we set the loss function at the boundary as the variance:
\begin{equation}\label{eq:bc_loss}
    \mathcal{L}_\text{BC} = \langle (u(\bm{r}, t) - u_0(\bm{r}, t))^2 \rangle_B,
\end{equation}
where $\langle \cdot \rangle_B$ denotes averaging over all points of $\bm{r} \in B$. The smaller this loss function is, the better the neural network satisfies the boundary conditions of the problem.

Let us define $\mathcal{L}_\text{PDE}$. If we have a PDE of \ref{eq:abs_pde} and a neural network $u(\bm{r}, t)$, substituting $f \rightarrow u$ and computing the RMS error of the PDE, we obtain the loss function:
\begin{equation}\label{eq:pde_loss}
    \mathcal{L}_\text{PDE} = \langle (\mathcal{D}[u(\bm{r}, t); \lambda])^2 \rangle_\Omega,
\end{equation}
where $\langle \cdot \rangle_\Omega$ again denotes the averaging over all points from the domain of definition of the $\Omega$ problem. The smaller this loss function is, the better the neural network satisfies the differential equations of the problem.

These two loss functions do not carry any information about the present exact solution (if it exists at all). Thus, the neural network is trained based on the given boundary conditions and physical laws (differential equations). That is why this approach to training is termed physics-driven.

\section{Data-driven approximation of the exact solution}\label{sec:data_driven}
In order to understand whether it is possible for a neural network model to learn the exact solution in physics-driven way, we can first solve a simpler data-driven problem. 

Let the exact solution functions $u(x, y), v(x, y), p(x, y)$ be known and set a grid of finite number of points $(x, y) \in \Omega$ in which these functions can be computed. We need to use a neural network to approximate the exact solution functions.
\subsection{Multi-stream Hybrid Network}\label{sec:qmodel}
\begin{figure*}[h]
    \centering
    \includegraphics[width=\textwidth]{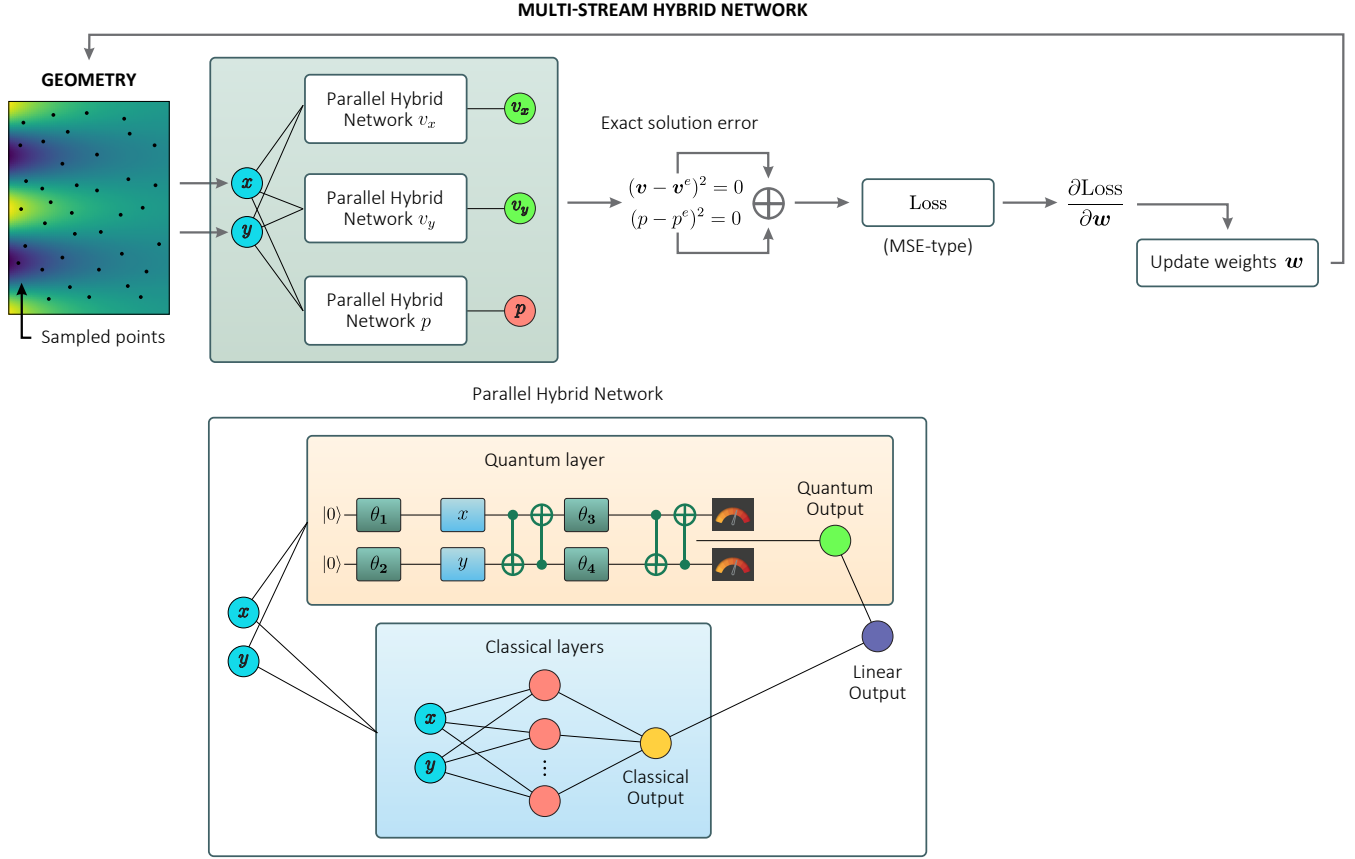}
    \caption{
    Overview of the Multi-stream Hybrid Network architecture. Spatial coordinates $(x, y)$ act as as input of three identical, but separate PHN layers. Each of these layers is responsible for predicting one component of the solution vector $(v_x, v_y, p)$. This is a data-driven model, so once the solution is predicted, the error (MSE) between exact and predicted solutions is calculated and used to update model weights at each training iteration.
    PHN layer architecture: input coordinates $(x, y)$ are passed through two parallel layers: quantum and classical. Classical layer is a 1-hidden-layer MLP. Quantum layer is a 
    parameterized two qubit circuit. The $R_X$ gates describe rotations about the $X$-axis on the Bloch sphere and are parameterized by the incoming coordinates, the $\text{Rot}(\bm \theta) = R_Z(\theta^1) R_Y(\theta^2) R_Z(\theta^3)$ gates describe arbitrary rotations and are parameterized by three trained weights $\bm{\theta}$. At the end of the circuit, both qubits are measured by the $\sigma_z$ operator. The measurement results of both qubits are passed through the ``Output'' layer to yield a single real value.
    }
    \label{fig:data_driven_architecture}
\end{figure*}

Here we introduce MHN -- Multi-stream Hybrid Network (without the ``physics'' part). MHN model is divided into 3 independent parts (PHNs), respectively for predicting $u, v$ and $p$, refer to Fig.~\ref{fig:data_driven_architecture}. This is motivated by the fact that these functions describe different physical values and their scale can differ. Each of PHN layers accept the $(x, y)$ coordinates. Then the outputs of these layers, respectively ``Quantum Output'' (\texttt{Qout}) and ``Classical Output'' (\texttt{Cout}) are affinely transformed, with an addition of cross-term $\texttt{Qout} \cdot \texttt{Cout}$
\begin{multline}
    \texttt{Output} = w_0 + w_1 \cdot \texttt{Qout} + w_2 \cdot \texttt{Cout} + w_3 \cdot \texttt{Qout} \cdot \texttt{Cout},
\end{multline}\label{eq:output}
where $w_i$ are the parameters to be trained. With the \texttt{Output} layer, it is as if we are giving the neural network a way to bring together the results of the quantum and classical parts of the network. The analysis of the parameter values after training is indeed consistent with this interpretation (the quantum part is usually responsible for approximating the periodic part of the solution, while the classical part contains attenuation and linear shift). The resulting quantity \texttt{Output} and is one of the predicted scalar quantities $(v_x, v_y, p)$. 

The classical layer is a small fully connected neural network with $h = 10$ neurons in the hidden layer, with activation functions \texttt{ReLU} standing between the layers. The quantum layer is a parameterized quantum two-qubit circuit, shown in Fig.~\ref{fig:data_driven_architecture}. The use of parameterized quantum circuits is widespread in quantum machine learning because it is a synthesis of the best ideas of classical machine learning and quantum computing~\cite{zhao2019qdnn, sagingalieva2023hybrid, dou2021unsupervised, kordzanganeh2023exponentially, haboury2024information, patapovich2025superposed}.

The chosen quantum circuit is quite simple, but demonstrates high efficiency. Training of quantum circuits on a simulator~\cite{kuzmin2025tqml} (classical computer) usually takes quite a long time as it is impossible to use backpropagation~\cite{rumelhart1986learning} when calculating gradients, so simple quantum circuits are trained faster. Current NISQ (noisy intermediate-scale quantum) devices, i.e., real quantum processors used today, are also not yet capable of high-precision computations on deep quantum circuits, so simplicity is a big advantage for us. Moreover, using a large number of qubits or a large depth in variational quantum circuits often leads to damped gradients -- Barren plateaus~\cite{mcclean2018barren, cerezo2021cost}.

We have defined all layers that make up MHN. The total number of parameters of this model is $936$. Training data will be generated as follows: on the problem domain $\Omega = [-0.5, 1.0] \times [-0.5, 1.5]$ we set a uniform grid of $n = 30 \times 40 = 1200$ points on which the model will make predictions and compare them with the exact solution \ref{eq:exact}. As a loss function, we choose the MSE error between the model prediction and the exact solution for each of the values $(v_x, v_y, p)$, and then add them up:
\begin{equation}
    \mathcal{L} = \mathcal{L}_{v_x} + \mathcal{L}_{v_y} + \mathcal{L}_{p},
\end{equation}
\begin{align}\label{eq:data_driven_loss}
\begin{split}
    \mathcal{L}_{v_x} &= \frac{1}{n} \sum_{i=1}^n (v_x^\text{p}(\bm{r}_i) - v_x^\text{e}(\bm{r}_i))^2, \\
    \mathcal{L}_{v_y} &= \frac{1}{n} \sum_{i=1}^n (v_y^\text{p}(\bm{r}_i) - v_y^\text{e}(\bm{r}_i))^2,\\
    \mathcal{L}_{p} &= \frac{1}{n} \sum_{i=1}^n (p^\text{p}(\bm{r}_i) - p^\text{e}(\bm{r}_i))^2, \\
\end{split}
\end{align}

\subsection{Multi-stream Classical Network}\label{sec:cmodel}
Let us define the MCN model -- a classical model which will be compared with MHN. For this purpose, let us take the architecture from Fig.~\ref{fig:data_driven_architecture} and replace quantum layer with the same classical one. We leave the activation function \texttt{ReLU} unchanged, as well as the PHN output function \ref{eq:output}. We choose the number of neurons in the hidden layer $h$ for both classical layers so that the number of parameters in MCN is greater than or equal to the total number of parameters in MHN. At $h = 12$, we obtain the number of parameters equal to $1239$. Recall that there were only $936$ parameters in MHN.

\subsection{Training}
Let us train both neural network models. In order to conduct training by the gradient descent method, we need to choose an optimizer program that will perform steps in the direction of loss function decrease, as well as update the model parameters. As mentioned before, we will choose the Adam algorithm as the optimizer.

\begin{figure}[H]
    \centering
    \includegraphics[width=1\linewidth]{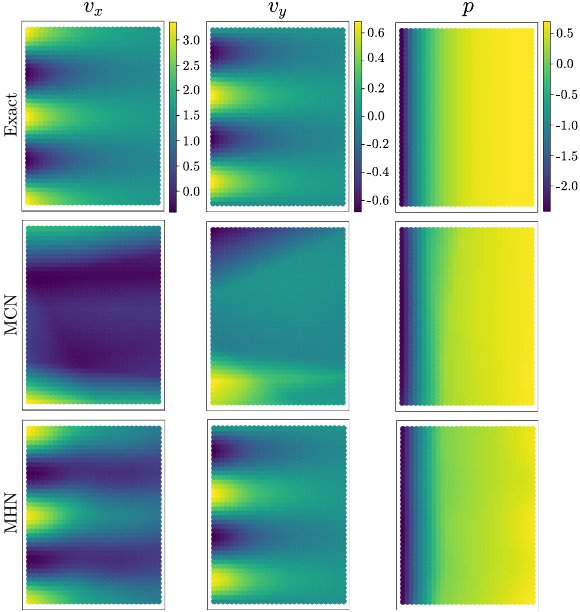}
    \caption{Training results of multi-stream network models after 100 epochs in data-driven learning. Here Exact, MCN and MHN denote the exact solution, classical multi-stream and hybrid multi-stream networks respectively, and $v_x, v_y, p$ are velocity and pressure projections. The MCN model does well in predicting the $p$ function, but cannot approximate the periodic functions $v_x$ and $v_y$. The MHN model approximates all solution functions equally well, with even fewer trainable parameters.}
    \label{fig:data_driven_comparison}
\end{figure}

As a result of training both models with the same learning rate $\alpha = 10^{-2}$ (selected as optimal as a result of comparing the results of training with different $\alpha$) for 100 epochs, we get the following results in Fig.~\ref{fig:data_driven_comparison}. We also calculate the root mean square error (RMSE) between the solutions of both models and the exact solution in Table~\ref{tab:data_driven_rmse}.

From these results, we can conclude that MHN has sufficient expressivity (the ability to approximate a wide class of functions) to learn the exact solution. In contrast, the classical model is unable to reproduce the exact solution, while having more trainable parameters. However, learning in the physics-driven approach uses a loss function completely different from the current one (defined in \ref{eq:data_driven_loss}), which has no information about the exact solution and relies only on a physical law -- the Navier-Stokes equation. We can say that we have established that proposed MHN satisfies the necessary condition for convergence to an exact solution, but does not necessarily satisfy the sufficient condition. We will clarify it in the next chapter.

\begin{table}[H]
    \centering
    \begin{tabular}{|l|l|l|}
    \hline
               & MHN       & MCN         \\ \hline
        $v_x$  & $0.3128$  & $0.7152$    \\
        $v_y$  & $0.0078$  & $0.1716$    \\
        $p$    & $0.0747$  & $0.0564$    \\ \hline
    \end{tabular}
    \caption{RMSE error between MHN and MCN models and the exact solution for $v_x, v_y, p$.}
    \label{tab:data_driven_rmse}
\end{table}

\section{Physics-driven training}
Here we will be dealing with physics-informed training of the MPHN model. Before we start training, let us make sure that the problem with the loss function defined in \ref{eq:physics_driven_loss} is actually solvable by neural networks. For this purpose, let us define a classical feedforward neural network (FNN) with a known large total number of parameters and try to train it.

\subsection{Feedforward Neural Network}\label{sec:fnn}
This FNN is a simple multilayer perceptron with $4$ hidden layers and \texttt{Tanh} activation functions, which takes two $(x, y)$ coordinates as inputs and yields three $(v_x, v_y, p)$ solution components.

Now let us define a strategy for selecting points inside our geometry and on its boundary (the loss functor $\mathcal{L}_\text{PDE}$ acts inside the geometry and $\mathcal{L}_\text{BC}$ acts on the boundary). We take a uniform distribution on our rectangular region $\Omega$ and use it to select $n_\text{PDE} = 2601$ inside the region and $n_\text{BC} = 400$ points on the boundary. In order to monitor the quality of the model's predictions during training, we choose a uniform grid of $n_\text{test} = 5000$ points (it includes both points inside the region and on its boundary), and compute the RMSE values for $v_x, v_y$ and $p$ separately every few epochs between the predictions and the exact solution \ref{eq:exact}.

As an optimizer, we again take the well-proven Adam algorithm, this time with the parameter $\alpha = 10^{-3}$ and train the FNN model for $1000$ epochs. As a result of training, the model reproduced the analytical solution quite accurately (see Fig.~\ref{fig:physics_driven_comparison}) with RMSE errors $v_x, v_y, p$ equal to $0.1249, 0.0468, 0.1162$, respectively. It would be possible to continue training and reduce the errors for all three values to the order of $10^{-5}$ (using, for example, a more resource-intensive optimizer L-BFGS~\cite{liu1989limited}), but this result is sufficient to demonstrate the success of a classical fully connected neural network.

\subsection{Multi-stream Physics 
Hybrid Network}\label{sec:physics_driven_qmodel}
\begin{figure*}[h!]
    \centering
    \includegraphics[width=\textwidth]{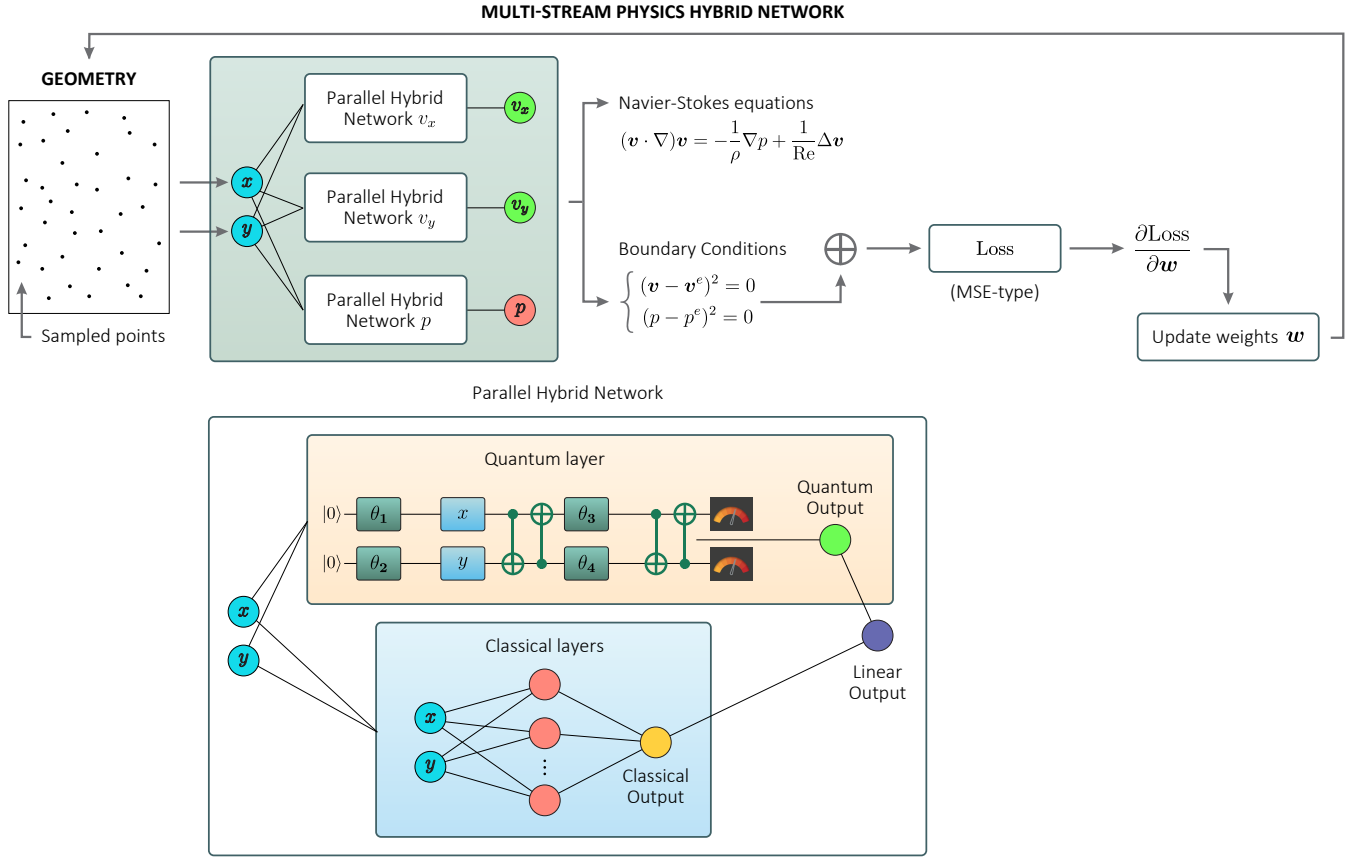}
    \caption{
    Overview of the MPHN architecture. It is identical to data-driven architecture in Fig.~\ref{fig:data_driven_architecture}. The only difference is in the loss function used for training. In MPHN, PDE and boundary conditions residuals are used as a loss, the model does not have any knowledge about the analytical solution.
    PHN layers architecture stays exactly the same.
    }
    \label{fig:physics_driven_architecture}
\end{figure*}

We now use the MPHN model to train on the physical loss function (Fig.~\ref{fig:physics_driven_architecture}). However, this time we replace the activation function \texttt{ReLU} by \texttt{SiLU}.

The advantage of \texttt{SiLU} over \texttt{ReLU} is that \texttt{SiLU} has non-zero smooth second and third derivatives, while already the second derivative of \texttt{ReLU} is identically zero. This fact is important because the Navier-Stokes equations \ref{eq:ns_kovasznay} contain the second derivatives of the velocities in coordinates, and the physical loss function $\mathcal{L}_\text{PDE}$ \ref{eq:pde_loss}, in turn, contains these very equations. Moreover, during the first-order optimization with the Adam algorithm, we will be forced to compute the gradients of the loss function over the model parameters, which will result in another derivative, this time not on the input data $(x, y)$, but on the trained weights $\bm{\theta}$. Therefore, using \texttt{ReLU} as the activation function can lead to damped gradients, which in turn leads to undertraining of the model and poor quality of predictions.

\begin{figure}[H]
    \centering
    \includegraphics[width=1\linewidth]{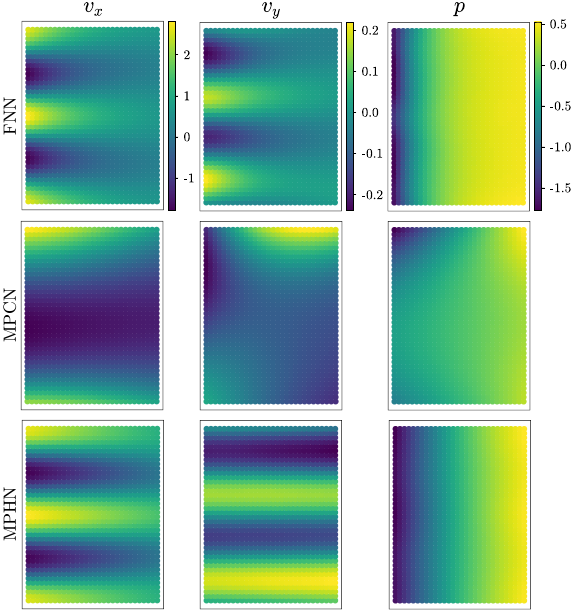}
    \caption{Training results of the FNN, MPCN, and MPHN models after 1000 epochs in physics-driven learning. Here $v_x, v_y, p$ are velocity and pressure projections. The FNN model, with a large number of parameters, was able to reproduce the exact solution with good accuracy. MPHN was able to learn the periodic nature of the solution for $v_x, v_y$, but has problems with the velocity decaying to zero along the $x$ axis. The MPCN did not remotely succeed in predicting the velocities. All networks were able to learn the pressure $p$ distribution quite well.}
    \label{fig:physics_driven_comparison}
\end{figure}

The number of points for training and testing, as well as the learning rate, is left the same as in FNN network ~\ref{sec:fnn}. The resulting velocity and pressure distributions after $1000$ epochs of training are shown in Fig.~\ref{fig:physics_driven_comparison}.

\subsection{Multi-stream Physics 
Classical Network}\label{{sec:physics_driven_cmodel}}
Let us introduce a classical variant of MPHN model -- Multi-stream Physics Classical Network or MPCN. We will apply this model to solve the physics-driven Kovasznay flow problem and compare the results with the previous Sec.~\ref{sec:physics_driven_qmodel}.  Recall that the MPCN differs from the MPHN simply by replacing the quantum layer containing the two-qubit quantum circuit with a copy of classical layer. The number of parameters of the classical model out numbers the hybrid one by a factor of $1.3$. All training parameters remain the same as in FNN and MPHN. The training results are shown in Fig.~\ref{fig:physics_driven_comparison} and RMSE errors are in the Table \ref{tab:physics_driven_rmse}.

\begin{table}[H]
    \centering
    \begin{tabular}{|l|l|l|l|}
    \hline
          & FNN    & MPCN   & MPHN   \\ \hline
    $v_x$ & 0.1249 & 0.7736 & 0.6308 \\ \hline
    $v_y$ & 0.0468 & 0.2245 & 0.1438 \\ \hline
    $p$   & 0.1162 & 0.7807 & 0.4588 \\ \hline
    \end{tabular}
    \caption{RMSE error between FNN, MPCN and MPHN models and the exact solution for $v_x, v_y, p$.}
    \label{tab:physics_driven_rmse}
\end{table}

\subsection{Results of physics-driven training}
As a result, we can state that the assumption we made for data-driven learning in Ch.~\ref{sec:data_driven} turned out to be correct: MPHN was indeed able to reproduce the exact solution in physics-driven learning setting, albeit with some error. In contrast, the MPCN was unable to capture any pattern of the exact solution. Even the simple exponential decay of pressure in \ref{eq:exact}, which the model captured in data-driven learning, could not be learned in physics-driven learning.
Also, on the example with data-driven learning, we can see that quantum circuits can approximate periodic functions better than classical ones, and thus can be thought as an analog of the ``natural'' Fourier transform in neural networks, which can find extensive application in many machine learning tasks (computer vision, sound analysis, etc.) \cite{schuld2021effect}.

\section{Discussion}
In this work, we proposed MPHN -- a new PINN-like architecture for physics modelling and evaluated it on the ``Kovasznay flow'' problem, which describes the flow of a fluid behind two-dimensional grid. 

The data-driven experiment showed that MHN is able to reproduce the exact solution with good accuracy, can capture its periodic and damped character. The classical analogue of this model, however, could not learn the exact solution; the periodic character of velocity projections $v_x, v_y$ was beyond the expressive capabilities of the MCN model.

Physics-driven experiment on the physical loss function confirmed the assumptions of the data-driven experiment: MPHN was able to replicate the character of the exact solution with good accuracy, without any information about the exact solution, only the Navier-Stokes equations and boundary conditions. The classical MPCN model, which was superior to MPHN in the number of parameters, did not succeed under similar conditions: even the damped character of the pressure, demonstrated in the data-driven experiment, was not learned by it. The large classical fully-connected FNN model, however, demonstrated high accuracy of the predictions of the exact solution. The obvious conclusion from this observation is that both hybrid and classical models will improve the quality of their predictions as the number of parameters increases. 

Thus, the results of this study demonstrate that for a comparable total number of parameters, with the same architecture, MPHN outperforms the MPCN (in terms of prediction accuracy), which may indicate that the model with a quantum circuit of just $2$ qubits is more expressive. When the depth and number of qubits of the quantum circuit are increased, we can expect a noticeable improvement in the quality of MPHN.

In conclusion, it is worth noting that there are many other neural network methods for solving differential equations, such as neural operators~\cite{neural_ops,fno}, which can learn immediately parameterized families of solutions, graph neural networks~\cite{mesh_based_gnn, sanchez2020learning}, which are capable of reversing the symmetries of the problem, as well as neural networks like \texttt{NeuralODE}~\cite{chen2018neural} with specially tailored architecture for solving ordinary differential equations.

\bibliography{lib}
\bibliographystyle{unsrt}

\end{document}